\documentstyle[12pt]{article}

\catcode `\@=11
\@addtoreset{equation}{section}

\catcode `\@=12

\newcommand{\be}{\begin{equation}}
\newcommand{\en}{\end{equation}}
\newcommand{\bea}{\begin{eqnarray}}
\newcommand{\ena}{\end{eqnarray}}
\newcommand{\beano}{\begin{eqnarray*}}
\newcommand{\enano}{\end{eqnarray*}}
\newcommand{\bee}{\begin{enumerate}}
\newcommand{\ene}{\end{enumerate}}
\newcommand{\Bbb}{\rm\bf}

\begin{document}

\thispagestyle{empty}

\vspace*{1cm}

\begin{center}
{\Large \bf Applications of Wavelets to Quantum Mechanics: a Pedagogical
Example }
\vspace{2cm}\\

{\large  F. Bagarello}
\vspace{3mm}\\
 Dipartimento di Matematica ed Applicazioni,
Facolt\'a di Ingegneria, Universit\`a di Palermo, I - 90128  Palermo, Italy.\\
E-mail:
Bagarello@Ipamat.Math.Unipa.It\\ \end{center}

\vspace*{2cm}

\begin{abstract}
\noindent
	We discuss in many details two quantum mechanical models of planar electrons
which are
very much related to the Fractional Quantum Hall Effect.

In particular, we discuss the localization properties of the trial ground
states of
the models starting from considerations on the numerical results on the energy.
We conclude
that wavelet theory can be conveniently used in the description of the system.

Finally we suggest applications of our results to the Fractional Quantum Hall
Effect.

\end{abstract}

\vfill

\newpage

\section{Introduction}

In these recent years a great effort has been done to
find a wave function which minimizes the energy of a
two-dimensional system of electrons subjected to a strong constant magnetic
field
 applied perpendicularly to the sample, independently of the electron density.
This is, in fact, the
first step to understand the main features of the Fractional Quantum Hall
Effect (FQHE). Many trial
ground states have been proposed so far, none of which has revealed to explain
all the
experimental data: the most successful is the one proposed by Laughlin,
\cite{lau} and
\cite{pra}, which describes an incompressible fluid (which therefore carries
current without
loosing energy) whose static energy is very low.

Totally different is the wave function proposed by Morchio, Strocchi and the
author in reference \cite{bag}. The authors, following the same line of other
authors,
\cite{lee}-\cite{maki}, consider the system of electron essentially as a
two-dimensional crystal. This crystal is built by first considering a
gaussian (a coherent state) centered at the origin, and then by "moving" this
gaussian
along the sites of a triangular lattice. The wave function of the finite volume
system is
the Slater determinant of the single electron wave functions centered in the
relevant lattice
sites. The details can be found in \cite{bag} where it is also shown that, for
low electron
densities, the energy of this state is lower than the one obtained by
Laughlin's state.
However the theoretical value of the $'$critical density$'$ at which the
crystal phase
appears favored with respect to the liquid one is slightly different from the
value given
by the experiments, see \cite{khurana}. Therefore, even if a crystal phase is
expected, its wave function must be refined.

In this paper we discuss a pedagogical model which suggests in which way one
can
modify the wave function in \cite{bag} to lower the energy, so to explain the
experimental data. The idea essentially consists in modifying the single
electron wave function trying to achieve a better electron localization. In
fact, we expect that the most localized the electron wave function is, the
lowest is the result obtained for the Coulomb energy if the electrons are
originally
localized around different spatial points. This claim follows both from
classical and
quantum considerations, see \cite{bonsall} and \cite{bag}.

The paper is organized as follows:

in Section 2 we introduce a physical model whose ground level is infinitely
degenerate (like the one of FQHE). In this way the ground state is not fixed a
priori. We
construct different trial ground states using the Haar, the
Littlewood-Paley and the harmonic oscillator bases. We also discuss their
localization properties.

In Section 3 we slightly modify the model previously introduced by fixing the
mean
positions of the electrons around lattice sites. Then we construct the new
basis and
we discuss how to compute the energies of the Coulomb interaction in these
different bases
for both the models considered.

In Section 4, we give and comment the numerical results.

In the Appendix, we introduce other models which can be treated with
analogous techniques, and in particular, we show that the FQHE belongs to this
class
of models.

\section{The model}

Let us consider a system of $N$ electrons living in a
two-dimensional device. We divide the hamiltonian in a single body
contribution plus a two-body term:
\be
H^{(N)}=\sum_{i=1}^N H_0(i) \hspace{3mm} +\frac{1}{2}\sum_{i\neq j}^N
\frac{1}{|{\bf r}_i-{\bf r}_j|}.
\label{hn}
\en
We observe that no background subtraction is considered in $H^{(N)}$. This is
meaningful only if we restrict to a finite number of electrons, that is, if we
keep $N$ to be
finite, otherwise the Coulomb energy diverges when $N \rightarrow \infty$.
Since we are not going to
compute the true energy of the system, but only a two-body contribution, we do
not need at this
stage to introduce the positive background.

The form of $H_0(i)$ is chosen here in a convenient way. In fact, we are more
interested in the
analogies of the model in (\ref{hn}) with the FQHE than to its physical
relevance.
Therefore, for reasons that will appear clear in the following, we take each
$H_0(i)$ of
the form \be
H_0=\frac{1}{2}(p_x^2+x^2)+\frac{1}{2}p_y^2+p_x p_y.
\label{h0}
\en
Therefore, each electron behaves like an harmonic oscillator in $x$, is free in
$y$, and is
also subjected to a $'$strange$'$ potential which is proportional to the
momentum of the
electron. We also see that the $z$-component does not appear in $H_0(i)$ (which
reflects the
fact that the device is two-dimensional).

It is easy to verify that the following canonical transformation
\bea
&Q&\equiv p_x+p_y; \hspace{20mm} P=-x; \nonumber \\
&Q'&\equiv p_y; \hspace{23mm} P'=x-y,
\label{tra}
\ena
preserves the commutation relations,
$$
[Q,P]=[Q',P']=i \hspace{1cm} [Q,P']=[Q,Q']=[P,P']=[P,Q']=0,
$$
and in this sense it is {\em canonical}, see \cite{mos}, and that in the new
variables $H_0$ take the form
\be
H_0=\frac{1}{2}(Q^2+P^2),
\label{osc}
\en
so that $Q'$ and $P'$ disappear from the definition of $H_0$.

To discuss the ground energy of the hamiltonian (\ref{hn}) we follow the same
steps as in
\cite{bag}: first, we find the ground state of the single electron unperturbed
hamiltonian
$H_0$. Then, we built up the trial ground state of the $N$-electrons
unperturbed
hamiltonian, $\sum_i H_0(i)$, as a Slater determinant of these single-electron
wave
functions. Finally, we compute the matrix element of the Coulomb interaction in
this state.
As widely discussed in \cite{bag}, this procedure seems to be justified at
least for small electron densities.

Before going on, let us show the relation between our pedagogical model
and the FQHE. The FQHE is described by the same hamiltonian as in
(\ref{hn}) with a different $H_0$,
$$
H_0^F=\frac{1}{2}(p_x-y/2)^2+\frac{1}{2}(p_y+x/2)^2,
$$
which describes an electron subjected to a constant magnetic field
perpendicular to the devise. A canonical transformation similar to the one in
(\ref{tra}), see \cite{bag} and references therein, transforms $H_0^F$ in the
same
$H_0$ (\ref{osc}). Therefore, even if $H_0$ in (\ref{h0}) and $H_0^F$ are
different, they
are both $'$projected$'$ into the same  harmonic oscillator (\ref{osc}) by
different
canonical transformations. This difference appears explicitly in the integral
transformation rule which relates the expression of the wave functions in the
variables
$(x,y)$ and $(Q,Q')$, see \cite{mos}. The reason why we discuss this
pedagogical example
and not directly the FQHE is that this transformation rule is very simple for
our $H_0$,
while it is much more difficult for $H_0^F$, see \cite{bag}. However, a first
real application of our
approach in the contest of FQHE can be found in \cite{ant}.

{}From \cite{mos} we can easily find that, if $\Psi(x,y)$ and $\Phi(Q,Q')$ are
respectively the wave functions in the ordinary space and in the
$'$canonically transformed$'$ space, they are related by
\be
\Psi(x,y) = \frac{1}{2 \pi}\int_{-\infty}^{\infty}dQ\,
\int_{-\infty}^{\infty}dQ' \, \Phi(Q,Q') \exp^{i[Q'(y-x)+Qx]}.
\label{tra2}
\en
To find out the single electron ground state of our model, $\Psi_0(x,y)$, it is
therefore
sufficient (actually equivalent) to find the ground state $\Phi_0(Q,Q')$ of
(\ref{osc}). Due
to the particular form of this $H_0$ we see that $\Phi_0(Q,Q')$ can be
factorized and that
the dependence on $Q$ is fixed:
regarding the Coulomb interaction as a perturbation of $H_0$, it is reasonable
to put
\be
\Phi(Q,Q')=\frac{1}{ \pi^{1/4}} \exp^{-Q^2/2} \, \phi(Q').
\label{fun}
\en
Here the function $\phi (Q')$ is totally free because the variable $Q'$ does
not
appear in the hamiltonian (\ref{osc}) and the energy of the unperturbed system
only depends
on $Q$. Only the perturbation fix its form. An analogous phenomenon, known
as the {\em degeneracy of the Landau levels}, takes place for $H_0^F$, for the
same reason. This explains why the ground state of the FQHE is not fixed.

We are now ready to discuss different choices of $\phi (Q')$ and
their projections in the configuration space via (\ref{tra2}), comparing the
resulting energies and localizations.

We start by considering the Littlewood-Paley orthonormal basis of wavelet. We
refer to \cite{dau} and \cite{chui} for all what concerns the wavelet
theory used in this paper.

The mother wavelet of this set is
$$
L(x) =(\pi x)^{-1}(\sin (2\pi x)-\sin (\pi x)),
$$
which, using the well known definition $L_{mn}(x)\equiv
2^{-m/2}L(2^{-m}x-n)$, generates an orthonormal set in $ L^2({\Bbb R})$. In
particular, we will be interested in the subset $\{L_m(x)\} \equiv
\{L_{m0}(x)\}$. The
functions of this set obviously satisfy the orthonormality condition
$$
<L_m,L_n>=\delta_{mn}
$$
and, when used in (\ref{tra2}) as different choices for the function $\phi
(Q')$, they
give the following set of wave functions in the configuration space:
\be
\Psi_m^{(LP)}(x,y)=\frac{2^{m/2}}{\sqrt{2} \pi^{3/4}} e^{-x^2/2} \chi_{{\em
D}_x}(y).
\label{lpw}
\en
Here $\chi_{{\em D}_x}(y)$ is the characteristic function of the set ${\em
D}_x$
, which is equal to one if $y \in {\em D}_x$ and zero otherwise.
We have defined
\be
{\em D}_x=[x-\frac{2\pi}{2^m},x-\frac{\pi}{2^m}]\cup [x+\frac{\pi}{2^m},x+
\frac{2\pi}{2^m}].
\label{set}
\en
Due to the canonicity of the transformation (\ref{tra}) the functions of the
set
$\{\Psi_m^{(LP)} (x,y): m \in \Bbb Z \}$ are obviously mutually orthonormal.
Moreover we see
from (\ref{lpw}) that they are very well localized in both $x$ and $y$.
\vspace{3mm}

Another possible choice for the function $\phi (Q')$ is any function belonging
to the
Haar wavelet set. The mother wavelet of this set is the function
$$
 H(x)=
  \left\{
  \begin{tabular}{rl}
  1, &  if $0\leq x <1/2$ \\
 $ -1,$ & if $1/2\leq x <1$ \\
  0, & otherwise
  \end{tabular}
  \right.
  $$
and the relevant set is defined in the usual way: $\{H_m(x)\} \equiv
\{H_{m0}(x)\} = \{ 2^{-m/2}H(2^{-m}x): m \in \Bbb Z  \}$. This set is again
orthonormal but the localization of each wave function is rather poor. From
(\ref{tra2}) we get
\be
\Psi_m^{(H)}(x,y)=\frac{2^{-m/2}i}{\sqrt{2} \pi^{3/4}} \frac{e^{-x^2/2}}{(y-x)}
(e^{i2^{m-1}(y-x)}-1)^2
\label{hw}
\en
which again decreases exponentially in $x$ but goes like $1/y$ in $y$. We do
not
expect therefore that the set $\{\Psi_m^{(H)} (x,y): m \in \Bbb Z \}$ can play
a
relevant role in the energy computation.

\vspace{3mm}
We end this section discussing another class of trial ground states of the
hamiltonian $H_0$. This time we will take non-wavelet functions. In particular,
we
consider for $\phi(Q')$ the first three eigenstates of the hamiltonian
$H_0=\frac{1}{2}(Q'^2+P'^2)$, which are orthonormal, and compute the
projections in the
configuration space of the complete function $\Phi(Q,Q')$ using the
transformation rule
(\ref{tra2}). We easily find the following results:
\bea
& \Psi_0^{(HO)}(x,y) & =\frac{1}{\sqrt{\pi}} \, e^{-(y^2+2x^2-2xy)/2} \nonumber
\\
& \Psi_1^{(HO)}(x,y) & =i\sqrt{\frac{2}{\pi}}\, (y-x) \, e^{-(y^2+2x^2-2xy)/2}
\label{how}  \\
& \Psi_2^{(HO)}(x,y) &=\frac{1}{\sqrt{2\pi}} \, (1-2(y-x)^2) \,
e^{-(y^2+2x^2-2xy)/2} \nonumber
\ena
All these wave functions, still mutually orthogonal, have a rather good
localization in
both $x$ and $y$. In particular the best localized is, of course,
$\Psi_0^{(HO)}(x,y)$.

The reason for considering also these wave functions in this paper is that
the choice of the ground state of the harmonic oscillator, which gives here
$\Psi_0^{(HO)}(x,y)$, is the one which allows the construction of the trial
ground state used in the description of the FQHE, see \cite{bag}. It is
therefore
useful, in our opinion, to have a comparison between these different
approaches.

\section{Energy Computation and Wavelet Bases on a Lattice}

We start this Section discussing the way in which the energy of the system can
be
computed. As a matter of fact, we are not going to compute the true energy of
the system, since in any case this is not physically very
interesting, but only a certain matrix element which is enough to get
some relevant informations on the ground state of the model, since it contains
the main contribution
to the energy. First of all, we fix $N=2$ in (\ref{hn}) since, in any case, the
total energy is
essentially a sum of two-body contributions.

In any book of Many-Body Theory it is shown that the computation of the
energy of an N-electrons system, in the Hartree-Fock approximation, is a
(summation of the) difference of two contributions, called respectively the
direct and the exchange terms. We should add to this difference also the ground
energy
of the kinetic Hamiltonian  $\sum_{i=1}^N H_0(i)$. However, in our model, as
well as in the
FQHE, this contribution is constant, in the sense that it does not depend on
the particular
choice of the function  $\phi$ in (\ref{fun}). Therefore it will be neglected
in all the
future considerations.

For our two-electrons system the wave function
is the following Slater determinant
\be
\Psi({\bf r}_1,{\bf r}_2)= \frac{1}{\sqrt{2!}} (\Psi_1({\bf r}_1)\Psi_2({\bf
r}_2)-\Psi_2({\bf r}_1)\Psi_1({\bf r}_2))
\label{sla}
\en
where $\Psi_i({\bf r}_j), \, i,j=1,2$, are the single electron wave functions
obtained in the previous Section, see (\ref{lpw}), (\ref{hw}) and (\ref{how}).
The Coulomb energy $E_c$ of the system is therefore
$$
E_c\equiv \int d^2{\bf r}_1 \,\int d^2{\bf r}_2 \, \frac{\Psi({\bf r}_1,{\bf
r}_2)^*
\Psi({\bf r}_1,{\bf r}_2)}{|{\bf r}_1-{\bf r}_2|}=V_d-V_{ex}
$$
where
$$
V_d= \int d^2{\bf r}_1 \,\int d^2{\bf r}_2 \, \frac{|\Psi_1({\bf r}_1)|^2
|\Psi_2({\bf r}_2)|^2}{|{\bf r}_1-{\bf r}_2|}
$$
and
$$
V_{ex}= \int d^2{\bf r}_1 \,\int d^2{\bf r}_2 \, \frac{\Psi_1({\bf r}_1)^*
\Psi_2({\bf r}_2)^* \Psi_1({\bf r}_2) \Psi_2({\bf r}_1)}{|{\bf r}_1-{\bf
r}_2|}
$$
are respectively the direct and the exchange term.

It is well known that,
at least for localized wave functions, the exchange contribution is much
smaller than the direct one. This feature is explicitly discussed, for
instance,
in \cite{bag}, where these contributions are explicitly computed for the FQHE.
Therefore in
this paper we will not compute $V_{ex}$, since it is not expected to change
significantly the
numerical results. We will focus our attention only on the computation of the
direct term $V_d$, which, with a little abuse of language, will be still often
called the $'$energy$'$ of the system.

Before computing the expressions of $V_d$ for the Littlewood-Paley and for
the harmonic oscillator bases we use these same bases as starting points to
introduce a $'$natural$'$ lattice in our model. The reason for doing this is
again
that a lattice is a natural structure for the FQHE at least for small electron
densities,
see \cite{bag}. Actually, since we are dealing with only two electrons, we will
think of our
lattice as two spatially not coincident points. We again refer to \cite{bag},
and reference
therein, for the details concerning the construction of the lattice associated
to $H_0$.

Using the results of the previous Section it is easy to prove that the
unitary operators
\be
T_1=e^{iQ'a} \hspace{2cm} T_2=e^{iP'b}
\label{shift}
\en
both commute with $H_0$ since they do not depend on $Q$ and $P$. Moreover, they
also commute
with each other if $ab=2 \pi N, \, \forall \, N\in {\Bbb Z}$.
{}From the
definition (\ref{tra}), one can also observe that, for any function $f(x,y) \in
L^2(\Bbb
R)$,
 $T_1f(x,y)=f(x,y+a)$ and $T_2f(x,y)=e^{i(x-y)b}f(x,y)$. Therefore, $T_2$ is
simply a
multiplication for a phase while $T_1$ acts like a shift operator. This is
enough for our
present aim: we can take the two different sites of our lattice along the
$y$-axis, with a $'$lattice$'$-distance $a=2\pi$.

In particular, defining
\be
\Phi_m^{(LP)}(x,y)\equiv T_1 \Psi_m^{(LP)}(x,y)=\frac{2^{m/2}}{\sqrt{2}
\pi^{3/4}} e^{-x^2/2} \chi_{{\em D}_{x-a}}(y)
\label{lpwl}
\en
from the Littlewood-Paley wavelets (\ref{lpw}), and
\be
\Phi_0^{(HO)}(x,y)\equiv T_1 \Psi_0^{(HO)}(x,y) =\frac{1}{\sqrt{\pi}}
e^{-((y+a)^2+2x^2-2x(y+a))/2}
\label{howl}
\en
for the most localized function of the harmonic oscillator states,
(\ref{how}), we conclude that both $\Phi_m^{(LP)}(x,y)$ and
$\Phi_0^{(HO)}(x,y)$ are
eigenstates of $H_0$ belonging to the ground level. This simply follows from
the
commutation rule $[T_1,H_0]=0$.

It is also easy to verify that $<\Psi_m^{(LP)},\Phi_m^{(LP)}>=0, \; \forall m
\geq 1$.

The
situation is a bit different for the oscillator wave functions; the
scalar product gives $<\Psi_0^{(HO)},\Phi_0^{(HO)}>=e^{-\pi^2}$. This implies
that the Slater
determinant is normalized within an error of $e^{-2\pi^2}=O(10^{-9})$.
Therefore this
extra contribution can safely be neglected here, and we will work with
$\Psi_0^{(HO)}$ and $\Phi_0^{(HO)}$ as if they where mutually orthogonal.
\vspace{3mm}

We continue this Section manipulating $V_d$ for two different models:

in the first one, which we
call the "Non-Lattice Model", the electrons are both localized around the
origin but
they are described by different wave functions (this is necessary in order not
to
annihilate the Slater determinant (\ref{sla}));

in the second one, the "Lattice Model", the electrons are
described by the same wave function localized around different space points. Of
course, this is the model which is more similar to the FQHE as already
discussed in \cite{bag}, and in this perspective it has a particular
interest.

We will omit the computation of the energy
with the Haar basis $\{ \Psi_m^{(H)}(x,y)\}$ since it is not expected to be
relevant for
understanding the FQHE. This is because the wave functions $\Psi_m^{(H)}(x,y)$
are the most
delocalized function within the ones we have introduced in the previous
Section, so that the Coulomb
energy is expected to be bigger than the one obtained by the other bases.

\subsection{Non-Lattice Model}

We start with manipulating the expression of $V_d$ for the basis in
(\ref{lpw}). From now on we will omit the index {\em d} since we will be
concerned only
with the direct contribution to the energy. Moreover, to explicitate the
dependence on the
quantum numbers $m,n$ and on the basis, we put
\be
V_{LP}^{m,n}=
\int d^2{\bf r}_1 \,\int d^2{\bf r}_2 \, \frac{|\Psi_m^{(LP)}({\bf r}_1)|^2
|\Psi_n^{(LP)}({\bf r}_2)|^2}{|{\bf r}_1-{\bf r}_2|}
\label{delp}
\en
where $m \neq n$ because of the Pauli principle. The integration in $y_1$ can
be easily
performed. After some manipulation and change of variables we can also perform
the
integration in $x_1$ and we obtain
\be
V_{LP}^{m,n}= \frac{2^{m+n}}{4\pi^3}
\sqrt{\frac{\pi}{2}}\int_{-\infty}^{\infty}\, dx e^{-x^2/2} \,
\int_{x+\pi/2^n}^{x+2\pi/2^n}\, dt \log {[\phi_{mn}(x,t)]}
\label{vdmn}
\en
where we have defined the following function
\beano
\phi_{mn}(x,t)&\equiv &
\frac{(t-\frac{\pi}{2^m}+\sqrt{x^2+(t-\frac{\pi}{2^m})^2})(t+\frac{2\pi}{2^m}
+\sqrt{x^2+(t+\frac{2\pi}{2^m})^2})}
{(t+\frac{\pi}{2^m}+\sqrt{x^2+(t+\frac{\pi}{2^m})^2})(t-\frac{2\pi}{2^m}
+\sqrt{x^2+(t-\frac{2\pi}{2^m})^2})} \times \\
& &\times
\frac{(t-\pi(\frac{3}{2^n}+\frac{1}{2^m})+\sqrt{x^2+(t-\pi(\frac{3}{2^n}+
\frac{1}{2^m}))^2})}{(t-\pi(\frac{3}{2^n}-\frac{1}{2^m})
+\sqrt{x^2+(t-\pi(\frac{3}{2^n}-\frac{1}{2^m}))^2})}\times\\
& &\times\frac{(t-\pi(\frac{3}{2^n}
-\frac{2}{2^m})+\sqrt{x^2+(t-\pi(\frac{3}{2^n} -
\frac{2}{2^m}))^2})}{(t-\pi(\frac{3}{2^n}
+\frac{2}{2^m}) +\sqrt{x^2+(t- \pi(\frac{3}{2^n}+\frac{2}{2^m}))^2})}
\enano
It is possible to see that the above integral is certainly defined for all $n$
different from $m\pm 1$. The integration can be easily performed numerically
and the results are discussed in Section 4.
\vspace{3mm}

To compute the energy for the harmonic oscillator wave functions (\ref{how})
it is better to use the following equality:
$$
 \frac{1}{|{\bf r}_1-{\bf r}_2|} = \frac{1}{2 \pi} \int \frac{d^2k}{|{\bf
k}|}\, e^{-i{\bf k}\cdot ({\bf r}_1-{\bf r}_2)}
$$
In this way the integrations in ${\bf r}_1$ and ${\bf r}_2$ in $V_d$ are
reduced to
gaussian integrals and therefore can be easily performed. Calling
$V_{ho}^{i,j}$ the $'$energies$'$ related to the wave functions in (\ref{how}),
we find:
\bea
&V_{ho}^{0,1}&= \frac{1}{2 \pi} \int \frac{d^2k}{|{\bf k}|}\,
(1-\frac{k_y^2}{2}) \, e^{-(k_x^2+2k_y^2+2k_xk_y)/2}\\
&V_{ho}^{0,2}&= \frac{1}{4 \pi} \int \frac{d^2k}{|{\bf k}|}\,
k_y^2 \cdot (2-\frac{k_y^2}{4}) \, e^{-(k_x^2+2k_y^2+2k_xk_y)/2}\\
&V_{ho}^{1,2}&= \frac{1}{4 \pi} \int \frac{d^2k}{|{\bf k}|}\,
(1-\frac{k_y^2}{2})\cdot k_y^2 \cdot (2-\frac{k_y^2}{4}) \,
e^{-(k_x^2+2k_y^2+2k_xk_y)/2}
\ena

\subsection{Lattice Model}

In this Subsection we use the wave functions  (\ref{lpwl}) and (\ref{howl})
obtained using
the shift operator $T_1$. We start considering the Littlewood-Paley basis.

As we have already said this time the energy is computed using the same
wave function centered in different lattice sites. Therefore it depends only on
a
quantum number, $m$.

We call $V_{LP}^{(m)}$ this energy
\be
V_{LP}^{(m)}=
\int d^2{\bf r}_1 \,\int d^2{\bf r}_2 \, \frac{|\Psi_m^{(LP)}({\bf r}_1)|^2
|\Phi_m^{(LP)}({\bf r}_2)|^2}{|{\bf r}_1-{\bf r}_2|}.
\label{delpl}
\en
The computation of this matrix element follows the same steps of the
calculus of the analogous contribution in (\ref{delp}), and one get a
similar expression:
\be
V_{LP}^{(m)}= \frac{2^{2m}}{4\pi^3}
\sqrt{\frac{\pi}{2}}\int_{-\infty}^{\infty}\, dx e^{-x^2/2} \,
\int_{x+2\pi+\pi/2^m}^{x+2\pi+2\pi/2^m}\, dt \, \log {[\phi_{m}(x,t)]}
\label{vdm}
\en
where
\beano
\phi_{m}(x,t)&\equiv &
\frac{(t+\frac{2\pi}{2^m}+\sqrt{x^2+(t+\frac{2\pi}{2^m})^2})
(t-\frac{4\pi}{2^m}+\sqrt{x^2+(t-\frac{4\pi}{2^m})^2})}
{(t+\frac{\pi}{2^m}+\sqrt{x^2+(t+\frac{\pi}{2^m})^2})
(t-\frac{5\pi}{2^m}+\sqrt{x^2+(t-\frac{5\pi} {2^m})^2})} \times \\
& &\times \frac{(t-\frac{\pi}{2^m}+\sqrt{x^2+(t-\frac{\pi}{2^m})^2})^2}
{(t-\frac{2\pi}{2^m} +\sqrt{x^2+(t-\frac{2\pi}{2^m})^2})^2}.
\enano
It is possible to prove that the integral surely exists for $m>1$, which is a
constraint
satisfied in our conditions since we are interested in studying the behavior of
the wave
functions and of the energy for large values of $m$. The reason of this
interest is that for
big $m$ the wave functions are more localized even in the variable $y$, as one
can see from
(\ref{lpw}) and (\ref{set}).

We see that the
result is very similar to the one for $V_{LP}^{m,n}$, as expected. However we
will show in
Section 4 that the numerical outputs are very different.

We end this Section simplifying the expression of the matrix element of the
Coulomb energy within the ground state of the harmonic oscillator and its
translated.
Using the integral formula for the Coulomb potential, we get
\bea
&V_{ho}^{0} &\equiv \int d^2{\bf r}_1 \,\int d^2{\bf r}_2 \,
\frac{|\Psi_0^{(HO)}({\bf
r}_1)|^2 |\Phi_0^{(HO)}({\bf r}_2)|^2}{|{\bf r}_1-{\bf r}_2|} =  \nonumber  \\
&= &\frac{1}{2\pi}\int\frac{d^2k} {|{\bf k}|}\, e^{-2\pi ik_y
-k_y^2-k_x^2/2-k_x k_y}
\label{vho}
\ena
The reason why only $\Psi_0^{(HO)}({\bf r})$ is considered here is essentially
that
this is the most localized function among all the harmonic oscillator wave
functions.

\section{Numerical Results and Comments}

In this Section we will discuss the numerical results for both the models
proposed previously
and we comment these results, giving particular attention to the localization
properties of
the wave functions.

We start by considering the Non-Lattice model. We report in Table 1 the
results for $V_{LP}^{m,n}$ for various values of $(m,n)$.

The energies of the harmonic oscillator are easily computed:
\be
V_{ho}^{0,1}=0.91873; \hspace{5mm} V_{ho}^{0,2}=0.39019;\hspace{5mm}
V_{ho}^{1,2}=0.11041
\en

\vspace{4mm}

For the Littlewood-Paley basis in the contest of the Lattice model the
situation is resumed in Table
2, while the energy of the harmonic oscillator is, in this case,
\be
V_{ho}^{0}=0.16515.
\en

\vspace{3mm}

We can now comment these results. The first obvious consideration is that,
while for the
first model the energy increases as much as $m$ and $n$ both increase, for the
lattice the
situation is just the opposite: the energy decreases for $m$ increasing. Let us
try to explain
this different behaviour. From the definitions (\ref{lpw}), (\ref{set}) and
(\ref{lpwl}) we see that
when $m$ increases the supports in $y$ of the functions decrease. Therefore
both
$\Psi_m^{(LP)}({\bf r})$ and $\Phi_m^{(LP)}({\bf r})$ improve their
localization for $m$
increasing. Since the electrons are localized at a distance of $2\pi$, we
return to a
situation similar to the one of the FQHE. It is well known that the lower bound
for the ground energy
is obtained if the electrons are punctually localized on the lattice sites,
that is, in the classical
limit. This is because in this way the distance between the electron is
maximized, and
therefore the Coulomb interaction gets its minimum value. These considerations
explain very
well the results in Table 2. We see, in fact, that when $m$ increases the
energy decreases
from $0.27083$ to the asymptotic value $0.16066$. This value is already reached
for $m=9$
and stay essentially unchanged even for bigger $m$. This value could also be
predicted in an
heuristic way. From the definition of $\Psi_m^{(LP)}({\bf r})$ we deduce that,
for $m$ very large,
this function behaves like a $\Psi_\infty^{(LP)}({\bf r})$ whose square modulus
is
\be
|\Psi_\infty^{(LP)}({\bf r})|^2 = \frac{1}{\sqrt{\pi}}\, e^{-x^2} \delta (x-y).
\label{psiin}
\en
An analogous formula holds for $\Phi_\infty^{(LP)}({\bf r})$,
$|\Phi_\infty^{(LP)}({\bf r})|^2 = \frac{1}{\sqrt{\pi}}\, e^{-x^2} \delta
(x-2\pi -y)$. We
can compute the energy $E_\infty$ in this limit and we get \beano
&E_\infty &\equiv \int d^2{\bf r}_1 \,\int d^2{\bf r}_2 \,
\frac{|\Psi_\infty^{(LP)}({\bf
r}_1)|^2 |\Phi_\infty^{(LP)}({\bf r}_2)|^2}{|{\bf r}_1-{\bf r}_2|} = \\
&=& \frac{1}{2\sqrt{\pi}}\int_{-\infty}^{\infty} \frac{dx \,
e^{-x^2/2}}{\sqrt{x^2+2\pi x+2
\pi^2}} = 0.16066
\enano
We observe that this result exactly coincides with the one obtained making $m$
increase. We
can conclude that the wave functions are really more and more localized since,
in fact,
their square modulus converges to an exponential function in $x$ times a
$\delta(x-y)$.

We furthermore observe that the energy of the harmonic oscillator,
$V_{ho}^{0}=0.16515$, is
slightly bigger than almost all the $V_{LP}^{(m)}$. This difference could be
interpreted
again as a better localization of the functions $\Psi_m^{(LP)}({\bf r})$ with
respect to
$\Psi_0^{(HO)}({\bf r})$. This is the reason why we have only computed the
energy for this
wave function and not, say, for $\Psi_1^{(HO)}({\bf r})$ or $\Psi_2^{(HO)}({\bf
r})$ whose
localization is a bit worse. These numerical results strongly suggest to use
wavelet
instead of oscillator functions for computing the energy even in the FQHE, in
the attempt of
 explaining better the phase transition between the Wigner and the Laughlin
phases, as discussed in the
Introduction.

Analogous conclusions can be obtained considering the results in Table 1. We
have left
these results for the end since they are less directly connected with the
picture of the
FQHE, since no lattice is present in this model.

We first add an extra information to Table 1: all the results turn out to be
symmetric
under the exchange $m\leftrightarrow n$, as they must.

This time the two electrons are localized both around the origin. We expect
that the
most localized the wave functions are, the maximum is the overlap between them
and,
therefore, the maximum is the energy. In a classical picture it would be like
if we put two
pointlike charges in the same point. Of course this system is not stable and we
expect a
very high energy for this configuration. This is exactly what happens. If we
try to compute
the energy $V_{LP}^{m,n}$ for $m$ and $n$ very large, we expect that the result
is the same
obtained by computing
$$
V_{LP}^{\infty}\equiv \int d^2{\bf r}_1 \,\int d^2{\bf r}_2 \,
\frac{|\Psi_\infty^{(LP)}({\bf
r}_1)|^2 |\Psi_\infty^{(LP)}({\bf r}_2)|^2}{|{\bf r}_1-{\bf r}_2|} =
\frac{1}{2\sqrt{\pi}}
\int_{-\infty}^{\infty} \frac{dx \, e^{-x^2/2}}{\sqrt{x^2}}
$$
which diverges, as expected.

Moreover, we see from Table 1 also that for each $m$ fixed, when $n$ increases,
$V_{LP}^{m,n}$
converges toward an asymptotic value. These values could be predicted with
great precision by
considering the following quantities:
\bea
&V_{LP}^{(m,\infty)}&\equiv \int d^2{\bf r}_1 \,\int d^2{\bf r}_2 \,
\frac{|\Psi_m^{(LP)}({\bf r}_1)|^2 |\Psi_\infty^{(LP)}({\bf r}_2)|^2}{|{\bf
r}_1-{\bf
r}_2|} = \nonumber \\
&=& \frac{2^{m-1}}{\pi^2}\sqrt{\frac{\pi}{2}} \int_{-\infty}^{\infty} dx \,
e^{-x^2/2} \log{[\Phi_m(x)]}
\label{44}
\ena
where we have defined
$$
\Phi_m(x)=\frac{(x+\frac{2\pi}{2^m}+\sqrt{x^2+(x+\frac{2\pi}{2^m})^2})
(x-\frac{\pi}{2^m}+\sqrt{x^2+(x-\frac{\pi}{2^m})^2})}
{(x+\frac{\pi}{2^m}+\sqrt{x^2+(x+\frac{\pi}{2^m})^2})
(x-\frac{2\pi}{2^m}+\sqrt{x^2+(x-\frac{2\pi}{2^m})^2})}
$$
The numerical results are reported in Table 3.
 We see that this results are extremely good, in the sense that they coincide
with the
asymptotic values of $V_{LP}^{m,n}$, for any $m$ fixed.

We observe also that, for any $m$ fixed, the energy decreases when $n$
increases. This can be
understood using the usual picture since, modifying only one wave function, the
overlap
between the two decreases.

Finally we observe again that the wavelets wave functions appear to be better
localized than the
oscillator ones. This is deduced, this time, since the best localized functions
correspond
to the maximum in energy. In fact, we have $V_{ho}^{0,1}=0.91873, \,
V_{ho}^{0,2}=0.39019$ and $V_{ho}^{1,2}=0.11041$. We see that the maximum of
these values
corresponds to the most localized wave function and, however, is much less than
the results
one gets using wavelets.

We conclude the analysis of these results again with the convintion that
wavelets can have a
strong utility in the problem of finding the ground state of the FQHE.
\vspace{50pt}

\noindent{\large \bf Acknowledgments} \vspace{5mm}

	It is a pleasure to thank Dr. R. Belledonne for her kind reading of the
manuscript.

\newpage

{\Large \bf {Appendix: Other Possible Models}}

\vspace{4mm}

In this Appendix we want to discuss briefly the main features of other physical
models which
could be treated with analogous techniques. We only discuss the situation in
two spatial
dimensions.

The essential ingredient to define the model is the single electron hamiltonian
$H_0$ in
(\ref{hn}). This operator must satisfy certain constraints. In fact it must be
so that it
exists a canonical transformation, generalizing the one in (\ref{tra}), which
transforms the
original $H_0$ in $x,y, p_x$ and $p_y$ in an hamiltonian depending only on a
couple
of conjugate variables. The most general linear transformation is the following
\beano
\tilde x_i = \sum_j (a_{ij}x_j+b_{ij}p_j) \\
\tilde p_i = \sum_j (c_{ij}x_j+d_{ij}p_j)
\enano
where $x_j$ and $p_j$ are the original canonically conjugate variables and
$\tilde x_i$ and
$\tilde p_i$ are the new ones. In \cite{mos} it is discussed this kind of
transformations and,
in particular, it is shown how the wave function transforms under this change
of variables.
In this paper, we have applied the results in \cite{mos} only to the
hamiltonian in
(\ref{h0}). Similar changes of variables can also be applied to other
hamiltonians, like the
following ones  \beano
&H_1&=\frac{1}{2}(p_x^2+x^2)+\frac{1}{2}y^2+x y  \\
&H_2&=\frac{1}{2}(p_x^2+p_y^2+x^2+y^2+2x y +2y p_x +(p_x x+x p_x))\\
&H_3&=\frac{1}{2}(p_x-y/2)^2+\frac{1}{2}(p_y+x/2)^2.
\enano
In particular, the last one is the Hamiltonian of the FQHE. All these
hamiltonians can be transformed
into the one of an harmonic oscillator with a suitable canonical
transformation. Of course the link
between the wave functions in configuration space and in the variables $(Q,Q')$
is different
depending on the coefficients $a_{ij},b_{ij},c_{ij}$ and $d_{ij}$ and it is
often not so
easy as in formula (\ref{tra2}).

The analysis of the FQHE will be discussed in a future paper.

\newpage

\vspace{3mm}

\begin{center}
{\large Table 1}
\end{center}

\vspace{2mm}

\begin{tabular}{|c||c||c|}
\hline
$V_{LP}^{(1,3)}=0.45784$           &$V_{LP}^{(1,4)}=0.43955$
&$V_{LP}^{(1,5)}=0.43520$ \\
\hline
$V_{LP}^{(1,6)}=0.43412$           &$V_{LP}^{(1,7)}=0.43385$
&$V_{LP}^{(1,8)}=0.43378$  \\
\hline
$V_{LP}^{(1,9)}=0.43377$         &$V_{LP}^{(1,m)}=0.43376 \; m\geq 10$
&  \\
\hline
$V_{LP}^{(2,4)}=0.77328$         &$V_{LP}^{(2,5)}=0.75524$
&$V_{LP}^{(2,6)}=0.75093$  \\
\hline
$V_{LP}^{(2,7)}=0.74986$         &$V_{LP}^{(2,8)}=0.74959$
&$V_{LP}^{(2,9)}=0.74953$  \\
\hline
$V_{LP}^{(2,10)}=0.74951$         &$V_{LP}^{(2,m)}=0.74950 \; m\geq 11$
&  \\
\hline
$V_{LP}^{(3,5)}=1.14544$         &$V_{LP}^{(3,6)}=1.12849$
&$V_{LP}^{(3,7)}=1.12445$  \\
\hline
$V_{LP}^{(3,8)}=1.12345$         &$V_{LP}^{(3,9)}=1.12320$
&$V_{LP}^{(3,10)}=1.12314$  \\
\hline
$V_{LP}^{(3,11)}=1.12313$         &$V_{LP}^{(3,m)}=1.12312 \; m\geq 15$
&   \\
\hline
$V_{LP}^{(4,7)}=1.51843$         &$V_{LP}^{(4,8)}=1.51449$
&$V_{LP}^{(4,9)}=1.51352$  \\
\hline
$V_{LP}^{(4,10)}=1.51327$         &$V_{LP}^{(4,11)}=1.51321$
&$V_{LP}^{(4,m)}=1.51319  \; m\geq 15$  \\
\hline
$V_{LP}^{(5,8)}=1.91054$         &$V_{LP}^{(5,9)}=1.90662$
&$V_{LP}^{(5,10)}=1.90565$  \\
\hline
$V_{LP}^{(5,11)}=1.90541$         &$V_{LP}^{(5,m)}=1.90533  \; m\geq 15$
&  \\
\hline
$V_{LP}^{(6,8)}=2.31870$         &$V_{LP}^{(6,9)}=2.30226$
&$V_{LP}^{(6,10)}=2.29835$  \\
\hline
$V_{LP}^{(6,11)}=2.29738$         &$V_{LP}^{(6,m)}=2.29705  \; m\geq 15$
&  \\
\hline
$V_{LP}^{(7,9)}=2.71004$         &$V_{LP}^{(7,10)}=2.69358$
&$V_{LP}^{(7,11)}=2.68966$  \\
\hline
$V_{LP}^{(7,m)}=2.68838 \; m\geq 15$         &     &  \\
\hline
$V_{LP}^{(8,10)}=3.10119$         &$V_{LP}^{(8,11)}=3.08475$
&$V_{LP}^{(8,15)}=3.07956$  \\
\hline
$V_{LP}^{(8,m)}=3.07954 \; m\geq 20$         &     &  \\
\hline
$V_{LP}^{(9,11)}=3.49229$         &$V_{LP}^{(9,12)}=3.47584$
&$V_{LP}^{(9,13)}=3.47192$  \\
\hline
$V_{LP}^{(9,15)}=3.47071$  &$V_{LP}^{(9,m)}=3.47063 \; m\geq 20$             &
\\
\hline
$V_{LP}^{(10,12)}=3.88337$         &$V_{LP}^{(10,13)}=3.86691$
&$V_{LP}^{(10,15)}=3.86202$  \\
\hline
$V_{LP}^{(10,m)}=3.86171 \; m\geq 20$         &     &  \\
\hline
$V_{LP}^{(15,20)}=5.81737$         &$V_{LP}^{(15,m)}=5.81705 \; m\geq 25$
&  \\
\hline
\end{tabular}

\vspace{4mm}

\indent
Table 1.-- Values of the matrix elements in (\ref{vdmn}) for different values
of $(m,n)$.

\newpage

\vspace{3mm}

\begin{center}
{\large Table 2}
\end{center}

\vspace{2mm}

\begin{tabular}{|c||c||c|}
\hline
$V_{LP}^{(1)}=0.27083$           &$V_{LP}^{(2)}=0.17462$
&$V_{LP}^{(3)}=0.16376$ \\
\hline
$V_{LP}^{(4)}=0.16141$           &$V_{LP}^{(5)}=0.16085$
&$V_{LP}^{(6)}=0.16071$  \\
\hline
$V_{LP}^{(7)}=0.16067$
&$V_{LP}^{(m)}=0.16066 \; m>7$      & \\
\hline
\end{tabular}

\vspace{4mm}

\indent
Table 2.-- Values of the matrix elements in (\ref{vdm}) for different values of
$m$.

\newpage

\vspace{3mm}

\begin{center}
{\large Table 3}
\end{center}

\vspace{2mm}

\begin{tabular}{|c||c||c|}
\hline
$V_{LP}^{(1,\infty)}=0.43376$           &$V_{LP}^{(2,\infty)}=0.74950$
&$V_{LP}^{(3,\infty)}=1.12312$ \\
\hline
$V_{LP}^{(4,\infty)}=1.51319$           &$V_{LP}^{(5,\infty)}=1.90533$
&$V_{LP}^{(6,\infty)}=2.29705$  \\
\hline
$V_{LP}^{(7,\infty)}=2.68838$         &$V_{LP}^{(8,\infty)}=3.07954$
&$V_{LP}^{(9,\infty)}=3.47063$  \\
\hline
$V_{LP}^{(10,\infty)}=3.86171$         &$V_{LP}^{(15,\infty)}=5.81704$
&$V_{LP}^{(20,\infty)}=7.77238$  \\
\hline
$V_{LP}^{(25,\infty)}=9.72772$         &$V_{LP}^{(30,\infty)}=11.68307$
&$V_{LP}^{(50,\infty)}=19.50437$  \\
\hline
$V_{LP}^{(100,\infty)}=39.05772$         &     &  \\
\hline
\end{tabular}

\vspace{4mm}

\indent
Table 3.-- Values of the matrix elements in (\ref{44}) for different values of
$m$.

\end{document}